\documentclass[pra,twocolumn,superscriptaddress,amsmath,amssymb,floatfix,noshowpacs]{revtex4}
\usepackage{dcolumn}
\usepackage{bm}
\usepackage{graphicx,hyperref}
\usepackage{color}
\usepackage{soul}
\usepackage{amsmath}
\usepackage{tabularx}
\graphicspath{{fig/}}
\usepackage{amsmath,amssymb}
\usepackage{centernot}
\usepackage{stmaryrd}
\usepackage{natbib}

\newcommand*{\bra}[1]{\langle #1 \vert}
\newcommand*{\ket}[1]{\vert #1 \rangle}

\begin{document}
\title{\textbf{Quantum entangled states of a classically radiating macroscopic spin}}
\author{Ori Somech}
\affiliation{Department of Chemical \& Biological Physics, Weizmann Institute of Science, Rehovot 7610001, Israel}
\author{Ephraim Shahmoon}
\affiliation{Department of Chemical \& Biological Physics, Weizmann Institute of Science, Rehovot 7610001, Israel}
\date{\today}

\begin{abstract}
Entanglement constitutes a main feature that distinguishes quantum from classical physics and is a key resource of quantum technologies.
Here we show, however, that entanglement may also serve as the essential ingredient for the emergence of classical behavior in a composite nonlinear radiating system.
We consider the radiation from a macroscopic spin emitter, such as the collective radiation from an atomic ensemble. We introduce a new class of macroscopic spin states, the coherently radiating spin states (CRSS), defined as the asymptotic eigenstates of the SU(2) lowering operator. We find that a spin emitter in a CRSS radiates classical-like coherent light, although the CRSS itself is a quantum entangled state exhibiting spin squeezing. We further show that CRSS are naturally produced in Dicke superradiance and underlie the dissipative Dicke phase transition. Our CRSS theory thus provides new concepts for studying the quantum physics of radiation, with applications in current platforms involving collections of atoms or spins, their consideration in quantum technologies such as metrology and lasing, and the many-body theory of spin systems.
\end{abstract}
\maketitle

The classical limit of radiation, such as that emitted by an antenna or laser, is typically associated with coherent states of the quantized electromagnetic field \cite{mandel_wolf_1995,MW,SCU}. This raises the question of what constitutes a classical emitter. Namely, considering a dipole emitter described quantum mechanically, what is the quantum state of this dipole which radiates a coherent state of the field. In a linear system, where the dipole is described by an harmonic oscillator, a coherent state of this oscillator linearly transforms to produce a coherent state also of the field. So, the classical-like state of the dipole radiates classical-like fields. One may wonder whether this latter conclusion is general and equally applies to relevant nonlinear quantum systems.

In particular, consider a dipole described by an SU(2) spin-$j$ wherein nonlinearity is exhibited by a finite Hilbert space of $2j+1$ states. This system is of fundamental and practical importance as it is often used to model the radiation from a collection of atom-like emitters, as in superradiance \cite{GROSS1982301,Dicke,mandel_wolf_1995,EMAN}: $N$ two-level atoms comprise a macroscopic dipole with a pseudo-spin $j=N/2\gg 1$. Such collective radiation effects appear in prominent quantum platforms involving ensembles of atoms \cite{HAR,TOMs1,TOMs2,FLD} or artificial emitters \cite{MAJ}. Hence, they play a key role in various quantum phenomena and technologies, ranging from dissipative phase transitions \cite{KES,DRUMMOND1978160} and spin squeezing \cite{Alejandro,A.M.Rey,yelin,REY} to narrowband superradiant lasers \cite{HOL,TOM1} and the precision of optical lattice clocks \cite{REY,CHAYE,HEN}.

The classical behavior of a macroscopic SU(2) spin is typically associated with coherent spin states (CSS) \cite{atomiccoherentstates,spinsqueezingreview,ASA}. These states are characterized by an average spin of length $j$ along a certain direction and a minimal uncertainty circle perpendicular to it (Fig. 1a) \cite{DowlingWigner}. Moreover, for a spin-$j$ composed of $N=2j$ two-level atoms, the CSS contains no entanglement: it is a product state of all constituent atoms. However, the radiation emitted by a spin $j$ in a CSS is not that of a classical dipole: in fact, it can exhibit quantum correlations typical of a nonlinear system, as we show below. So, CSS are classical-like states of the spin itself but they radiate non-classical light.
\begin{figure}
    \centering
    \includegraphics[width=\columnwidth]{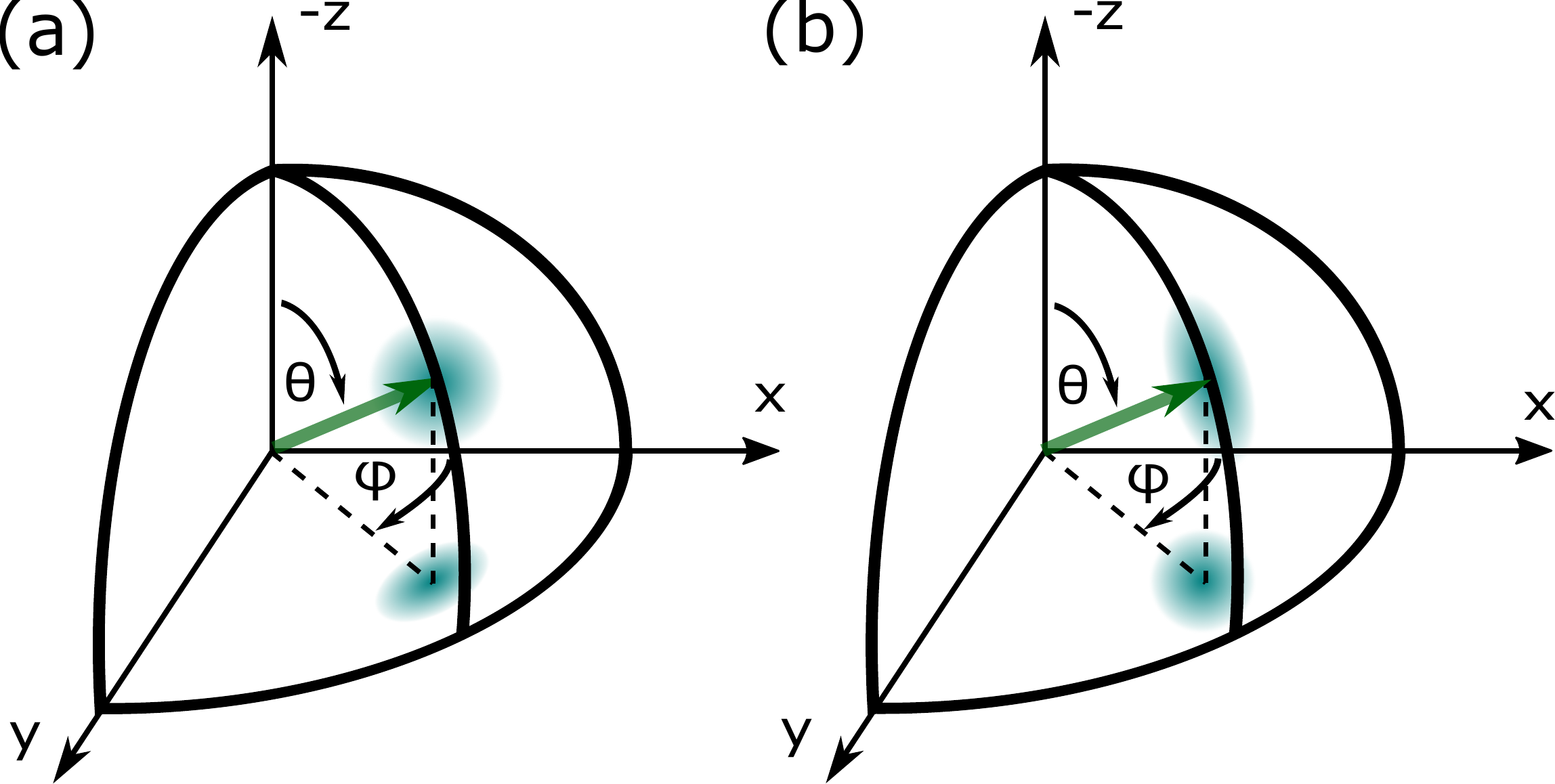}
    \caption{
    Spin states represented on the Bloch sphere (southern hemisphere). (a) CSS: the variance contour perpendicular to the mean spin vector (green arrow) exhibits no spin squeezing (circular contour). However, its projection onto the $xy$ ``dipole plane" \emph{is} squeezed (elliptical), leading to non-classical radiation (see \emph{Dipole-projected squeezing} below). (b) CRSS exhibits spin squeezing (elliptical contour perpendicular to mean spin) but no dipole-projected squeezing (circular contour on $xy$ plane), hence it radiates classical light.
    }
    \label{fig1}
\end{figure}

In contrast, here we introduce the \emph{coherently radiating spin state} (CRSS) as the state of a spin system which produces classical-like coherent state radiation (Fig. 1b). We find that such states exist in the macroscopic limit $j\gg 1$ and are given by the approximate eigenstates of the spin lowering operator $\hat{J}=\hat{J}_x-i\hat{J}_y$ with continuous complex eigenvalues $\alpha$ restricted to $|\alpha|<j$. This is in analogy to bosonic coherent states defined as the eigenstates of the non-Hermitian boson lowering operator, and in contrast to CSS which are eigenstates of an Hermitian spin operator. We show that the CRSS possesses several interesting features: (1) CRSS is a nonclassical state exhibiting spin squeezing which increases as $|\alpha|\rightarrow j$. For a spin-$j$ describing $N=2j$ two-level atoms this means that the constituent atoms are pairwise entangled \cite{LEW,SOR,spinsqueezingreview}. (2) CRSS is the steady-state of driven-dissipative Dicke superradiance \cite{DRUMMOND1978160,CAR,LAW,RABk}, and is at the origin of the dissipative Dicke phase transition \cite{DRUMMOND1978160,CAR,KES} and associated spin squeezing \cite{Alejandro,yelin,A.M.Rey}. This allows us to analytically find the scaling $N^{-\frac{1}{3}}$ of the best achievable spin squeezing. (3) The radiated light from a CRSS is a coherent state, whose amplitude is proportional to $\alpha$ and to the coherent drive amplitude, just as in a linear system.
CRSS thus forms a remarkable case of the quantum-classical border, wherein a composite quantum nonlinear system is perceived by light as a classical linear scatterer, provided that its constituents are quantum entangled.

\noindent\\
\textbf{Radiation from a dipole eigenstate}
\\
The idea that an eigenstate of the lowering operator $\hat{J}$ should produce coherent-state radiation is in fact quite generic and can be intuitively understood as follows. Consider a Hermitian dipole operator in the general form $\hat{d}_{\mathrm{H}}=\hat{d}+\hat{d}^{\dag}$ coupled to the Hermitian field $\hat{E}_{\mathrm{H}}=\hat{E}+\hat{E}^{\dag}$ (with $[\hat{E},\hat{E}^{\dag}]= 1$), both taken dimensionless here. The light-matter interaction Hamiltonian, $\propto\hat{d}_{\mathrm{H}}\hat{E}_{\mathrm{H}}$, within the rotating-wave approximation \cite{SCU}, takes the form
\begin{equation}
\hat{H}\propto\hat{d}\hat{E}^{\dag}+\hat{E}\hat{d}^{\dag}.
\label{H0}
\end{equation}
If the system is constantly pumped into an eigenstate $|\alpha\rangle$ of $\hat{d}$ with eigenvalue $\alpha$ then we may effectively replace $\hat{d}\rightarrow \alpha$ in Hamiltonian (\ref{H0}) such that the evolution $e^{-\frac{i}{\hbar} \hat{H} t}$ is the displacement operator of the field, generating a coherent-state out of an initial vacuum state (Appendix). So, an eigenstate of $\hat{d}$ produces coherent-state radiation. Indeed, for a linear dipole, where $\hat{d}$ is a boson lowering operator, $|\alpha\rangle$ is a coherent state of the dipole, which linearly transforms via (\ref{H0}) to produce a coherent-state field. However, for the nonlinear dipole represented by a spin $j$, we first consider the possible existence of eigenstates of $\hat{d}=\hat{J}$ and then study their generation and radiation properties.

\noindent\\
\textbf{Asymptotic eigenstates of $\hat{J}$}
\\
Strictly speaking, the operator $\hat{J}$ does not have any eigenstates except for the eigenstate $\ket{j,-j}$ with an eigenvalue zero, where we use the common notation $\ket{j,m}$ for the joint eigenstate of the total angular momentum operator $\hat{J}_x^2+\hat{J}_y^2+\hat{J}_z^2$ and its $z$-axis projection $\hat{J}_z$, with corresponding eigenvalues $j(j+1)$ and $m$. However, we show below that for a macroscopic spin, $j \gg 1$, there are states that are approximate eigenstates of $\hat{J}$ up to excellent accuracy. We call such states CRSS. Speaking loosely, we find that in the limit $j\rightarrow\infty$ the operator $\hat{J}$ has a set of eigenstates with complex eigenvalues $\alpha$ satisfying
\begin{equation}
\hat{J}\ket{j,\alpha}=\alpha\ket{j,\alpha}, \quad \alpha= jre^{-i\varphi}, \quad 0<r<1, \quad r,\varphi\in\mathbb{R}.
\label{eig}
\end{equation}
More formally, we define this condition as,
\begin{equation}
\text{\ensuremath{\underset{j\rightarrow\infty}{\text{lim}}}}\epsilon_{j}(r)=0, \quad
\epsilon_{j}(r)\equiv\left\Vert \hat{J}\ket{j,jre^{-i\varphi}}-jre^{-i\varphi}\ket{j,jre^{-i\varphi}}\right\Vert.
\label{limit}
\end{equation}
Here $\epsilon_{j}(r)$ is the proximity error of a state $\ket{j,\alpha}$ for being an eigenstate of the operator $\hat{J}$ with an eigenvalue $\alpha=jre^{-i\varphi}$ ($r<1$), and we request that this quantity approaches zero for $j\rightarrow\infty$. Importantly, the ratio $\alpha/j=re^{-i\varphi}$ is kept constant while taking the limit $j\rightarrow\infty$. This allows to account for excitation amplitudes $\alpha$ comparable to $j$, where nonlinearity becomes significant (beyond the Holstein-Primakoff approximation).

\begin{figure}
    \centering
    \includegraphics[width=\columnwidth]{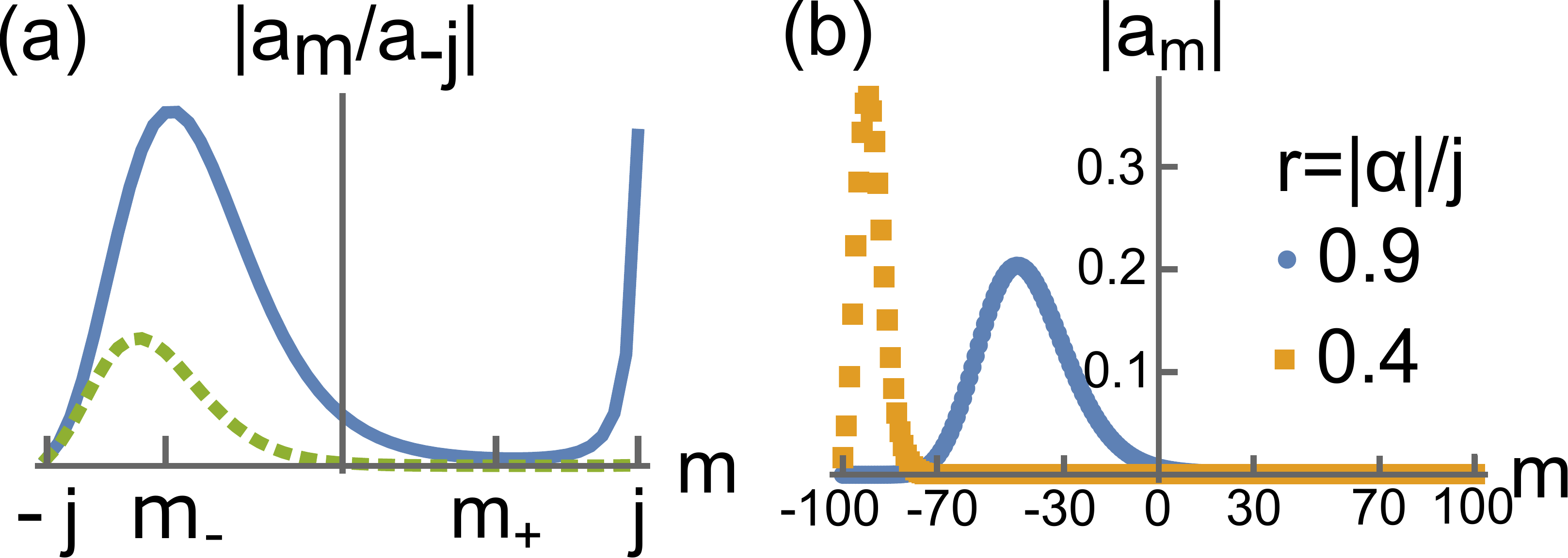}
    \caption{
    CRSS as approximate eigenstates of the spin-$j$ lowering operator $\hat{J}$.
    (a) State coefficients $|a_m/a_{-j}|$ from Eq. (\ref{coeff}) with $j=25$. For $r=0.75$ (dashed line) $a_{j}$ approximately vanishes whereas for $r=0.84$ (solid line) it does not. For the latter, the maximum and minimum points $m_{\mp}$ are marked. In the CRSS ansatz (\ref{ans}), all coefficients $m>m_+$ are set to zero, and the condition (\ref{limit}) is satisfied for all $r<1$. (b) Normalized coefficients $|a_m|$ of the ansatz state (\ref{ans}) for $j=100$. The Gaussian shape agrees well with the analytical result (\ref{am}).
    }
    \label{fig2}
\end{figure}

We show a simple way to define a state that satisfies Eq. (\ref{limit}).
Assuming Eq. (\ref{eig}) can be solved, we insert to the latter the general state $\sum_{m=-j}^j {a_m\ket{j,m}}$ and obtain the recursion relation,
\begin{equation}
a_{m+1}=\frac{\alpha}{\sqrt{j(j+1)-m(m+1)}}a_{m},\quad -j\leq m\leq j-1,
\label{recursion}
\end{equation}
together with $a_j=0$. These two conditions result in a contradiction for any $|\alpha|>0$ unless $a_j$ tends to zero in the required limit $j\rightarrow \infty$. The solution for the recursion relation (\ref{recursion}) is given by
\begin{eqnarray}
a_m&=&a_{-j} e^{f_m}e^{-i(m+j)\varphi},
\nonumber\\
f_m&=&(m+j)\ln(rj)-\frac{1}{2}\sum_{k=-j}^{m-1}\ln\left[j(j+1)-k(k+1)\right].
\label{coeff}
\end{eqnarray}
In Fig. \ref{fig2}a we plot $|a_m/a_{-j}|$ as function of $m$, observing that depending on $r$, $a_j$ does not always tend to zero. We find analytically and numerically that as $j\rightarrow \infty$, $a_j\rightarrow 0$ only for $r<0.804...$ \cite{SM}. To satisfy Eq. (\ref{limit}) in the full domain $0<r<1$, we use the following observation: for a truncated state $\ket{\psi_s}=\sum_{m=-j}^{s}a_m\ket{j,m}$ with $s<j$, we obtain that the proximity error is proportional to the last coefficient $\left\Vert \hat{J}\ket{\psi_s}-\alpha\ket{\psi_s}\right\Vert=|\alpha a_{s}|$. Using Eq. (\ref{recursion}), we can identify the regions for which $|a_m|$ is increasing or decreasing as a function of $m$, finding the maximum ($m_-$) and minimum ($m_+$) points as the roots of the quadratic equation $|\alpha|^{2}=j(j+1)-m(m+1)$, $m_{\pm}=\frac{1}{2}\left(-1\pm\sqrt{-4|\alpha|^{2}+4j^{2}+4j+1}\right)$ (rounded to integers), see Fig. \ref{fig2}a.
For the minimal proximity error we thus choose to truncate at $s=m_+$, defining our ansatz eigenstate and resulting error as
\begin{eqnarray}
|j,jre^{-i\varphi}\rangle_{\mathrm{ans}}=\sum_{m=-j}^{m_+}a_m\ket{j,m}, \quad \epsilon_{j}(r)=jr |a_{m_+}|.
\label{ans}
\end{eqnarray}
The error can be readily estimated numerically for given $j$ and $r$, using Eq. (\ref{coeff}) for $|a_{m_+}|$ and demanding state normalization $|a_{-j}|=\left[\sum_{m=-j}^{m_+}e^{2f_m}\right]^{-1/2}$. This is presented in Fig. 3, observing that for $j\gg 1$, the error decreases exponentially with $j$ for any $r<1$. We also find that at the limit $j\rightarrow\infty$ the state (\ref{ans}) minimizes the error $\epsilon_{j}(r)$ for fixed $j$ and $r$ (Appendix). We thus conclude that the state (\ref{ans}) is an eigenstate of $\hat{J}$ in the sense of Eq. (\ref{limit}), i.e. it is a CRSS to an excellent approximation.

It is instructive to estimate the error from Eq. (\ref{ans}) also analytically. To that end, we write
$|a_{m_+}|=e^{f_{m_+}-f_{m_-}}|a_{m_-}|$, approximate $f_{m_{\pm}}$ by converting sum to integral in Eq. (\ref{coeff}), and estimate $|a_{m_-}|$ using  (\ref{am}) below, obtaining for $j\gg 1$
\begin{eqnarray}
\epsilon_{j}(r)\approx q e^{-2j g(r)}, \: g(r)=\mathrm{arctanh}[\sqrt{1-r^2}]-\sqrt{1-r^2},
\label{eps}
\end{eqnarray}
with the power-law prefactor $q(j,r)=j^{\frac{3}{4}}(r^2\sqrt{1-r^2}/\pi))^{\frac{1}{4}}$.
Since $g(r)>0$ for $r<1$, the error (\ref{eps}) decays essentially exponentially with $j$, exhibiting excellent agreement with the numerical results of Fig. 3, and yielding the CRSS existence condition $r<1$. The exponent $g(r)$ is a monotonically decreasing function of $r$, meaning that for increasing values of $r$ larger $j$ values are required for the error to still be small, as also seen in Fig. 3. For a given finite $j\gg 1$, we can thus define the range of validity $0<r<r_{j}$ in which the state (\ref{ans}) is a CRSS to a good approximation, by defining the respective maximal value $r_j$ via
 \begin{subequations}
   \begin{tabularx}{\hsize}{@{}XXX@{}}
     \begin{equation}
        \epsilon_{j}(r=r_j)=\frac{1}{e},
        \label{rj1}
     \end{equation} &
     \begin{equation}
        r_j\sim \sqrt{1-\left(\frac{3}{2j}\right)^{\frac{2}{3}}},
        \label{rj2}
     \end{equation}
   \end{tabularx}
 \end{subequations}
as marked by the dashed curve in Fig. 3b. The analytical result (\ref{rj2}) for $r_j$ expresses its asymptotic scaling with $j$, which is simply obtained by demanding $2jg(r_{j})=1$ (ignoring the power-law prefactor $q$ for $j\rightarrow \infty$).

\begin{figure}
    \centering
    \includegraphics[width=\columnwidth]{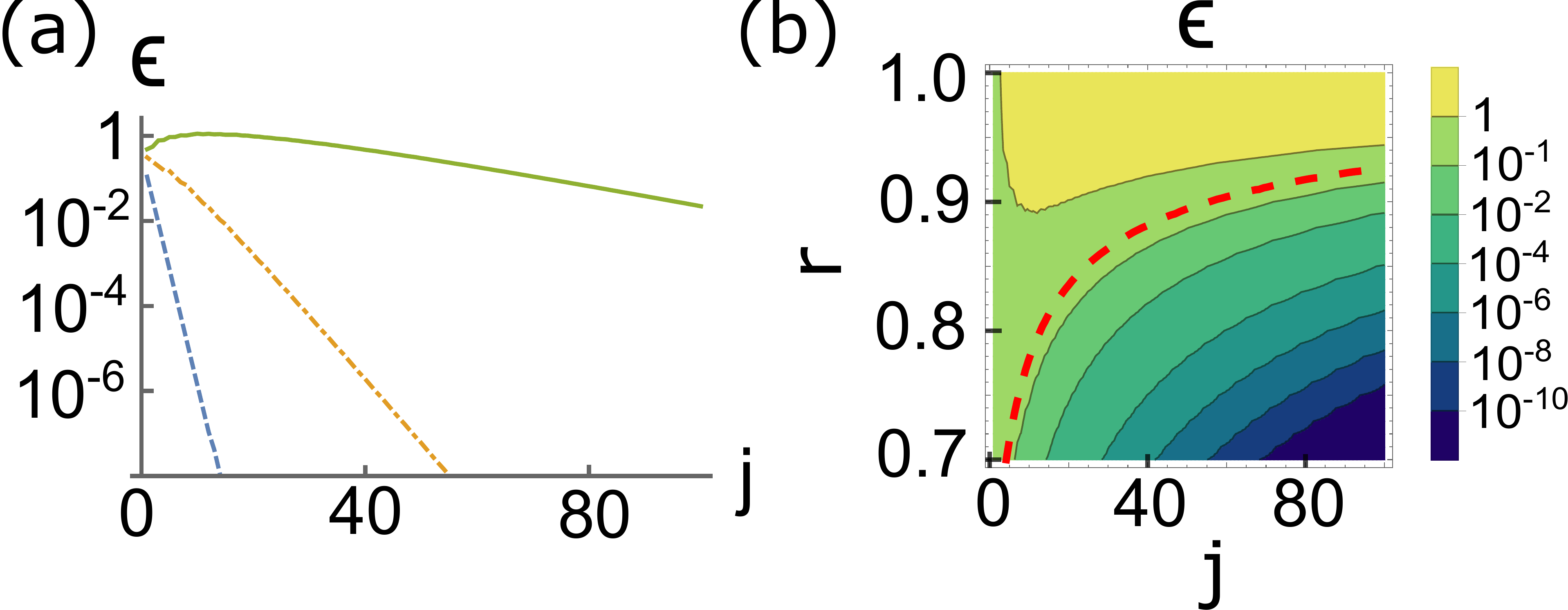}
    \caption{
    Proximity error of the CRSS ansatz state (\ref{ans}) [evaluated numerically using (\ref{coeff})]. (a) $\epsilon_{j}(r)$ vs. $j$ for $r=0.4,0.7,0.9$ (dashed, dash-dotted, solid lines). Exponential decay is observed for large $j$ in agreement with (\ref{eps}).
    (b) $\epsilon_{j}(r)$ as a function of $j$ and $r$. The dashed curve marks $r_j$, defined in (\ref{rj1}) and calculated using (\ref{eps}). For given $j$ and $0<r<r_j$, the state (\ref{ans}) well approximates a CRSS ($\epsilon\ll 1$), as predicted analytically.
    }
    \label{fig3}
\end{figure}

\noindent\\
\textbf{Steady state of driven superradiance}
\\
Consider a dipole $\hat{d}=\hat{J}$ damped by radiation to a continuum of electromagnetic field modes described by the operator $\hat{E}$. The master equation for the reduced density matrix of the dipole can be derived from Hamiltonian (\ref{H0}) as \cite{GROSS1982301,Alejandro},
\begin{eqnarray}
\frac{d\hat{\rho}}{d t}&=&-\frac{i}{\hbar}\left(\hat{H}_{\text {nh }} \hat{\rho}-\hat{\rho} \hat{H}_{\text {nh }}^{\dagger}\right)+\gamma \hat{J} \hat{\rho} \hat{J}^{\dagger},
\nonumber\\
\hat{H}_{\text {nh}}&=&\hbar\left(\Delta-i\frac{\gamma}{2}\right)\hat{J}^{\dag}\hat{J}-\hbar\left(\Omega \hat{J}^{\dag} +\Omega^{\ast} \hat{J}\right).
\label{ME}
\end{eqnarray}
Here $\gamma$ and $\Delta$ are spontaneous emission rate and energy shift, respectively, both induced by the electromagnetic field reservoir, and $\Omega$ is the amplitude of a resonant coherent-state incident field.
In the case of superradiance, $\hat{J}=\sum_{n=1}^N \hat{\sigma}_n$ describes the collective pseudo-spin of $N$ two-level atoms with lowering operators $\hat{\sigma}_n=|g\rangle_n\langle e|$ (levels $|g\rangle$ and $|e\rangle$), which are all identically coupled to the field $\hat{E}$, as can be realized in an optical cavity \cite{HAR,TOMs1}. The common field reservoir can then mediate correlations between the atoms via the collective decay ($\gamma$) and dipole-dipole interaction ($\Delta$). Assuming that all atoms are initially in the ground state, $\prod_n\otimes |g\rangle_n=|N/2,-N/2\rangle$, the spin representation is fixed to $j=N/2$ throughout the dynamics. The steady state of this Lindblad master equation becomes a pure state if and only if it is an eigenstate of both the non-Hermitian Hamiltonian $\hat{H}_\text{nh}$ and the jump operator $\hat{J}$ \cite{purestate1}. We showed that the eigenstate (\ref{eig}) of $\hat{J}$ exists at the limit $N\rightarrow \infty$ (CRSS) and it is straightforward to verify that it is also an eigenstate of $\hat{H}_\text{nh}$ if we take
\begin{eqnarray}
\alpha=\frac{\Omega}{\Delta-i\gamma/2},
\label{alp}
\end{eqnarray}
yielding $d\hat{\rho}/dt=0$ for $\hat{\rho}=\ket{j,\alpha}\bra{j,\alpha}$ with $j=N/2$. This can also be seen for a finite $N=2j\gg 1$ by considering the overlap between the exact steady-state solution of (\ref{ME}) and the CRSS ansatz (\ref{ans}), whose deviation from unity behaves similarly to the proximity error in Fig. 3 (Appendix).

The above considerations establish that for $N=2j \gg 1$, the steady state of driven-dissipative superradiance is a CRSS characterized by the amplitude $\alpha$ from Eq. (\ref{alp}), within the validity range $0<r<r_j$ defined in Eq. (\ref{rj1}) for a finite $j$. This means that the properties of CRSS, a pure state, underlie dissipative Dicke superradiance phenomena, as we show further below. For example, in the macroscopic limit $N=2j\rightarrow \infty$, the steady-state solution of Eq. (\ref{ME}) is known to exhibit a second order phase transition wherein the order parameter $\langle \hat{J}_z\rangle$ vanishes for driving fields $|\Omega|$ exceeding the critical point $\Omega_c =(N/4)\sqrt{\gamma^2+4\Delta^2}$ \cite{DRUMMOND1978160,A.M.Rey,HL}. But the condition $|\Omega|<\Omega_c$ for non-vanishing order parameter is identical to the condition $|\alpha|/j=r<1$ for the existence of CRSS in Eq. (\ref{eig}). Therefore, the dissipative Dicke phase transition is originated in the critical point $r=1$ for the existence of eigenstates of $\hat{J}$ in the limit $j\rightarrow \infty$.

\noindent\\
\textbf{Spin observables and squeezing}
\\
Having established the existence of CRSS and its relation to superradiance, we now examine the properties of spin variables in a CRSS. Using Eq. (\ref{eig}), the average dipole $\langle \hat{J} \rangle$ is trivially given by the amplitude $\alpha$, agreeing with mean-field results of superradiance \cite{DRUMMOND1978160,yelin,HL}. Notably, in the case of superradiance, Eq. (\ref{alp}) implies that this dipole amplitude is linear with the incident field $\Omega$, a result expected for a harmonic oscillator but less intuitive for our nonlinear spin scatterer. We note that this is valid for very strong fields $|\alpha|\sim j\gg 1$ well beyond the typical linear limit $|\Omega|\ll \Omega_c$ ($|\alpha|\ll j$). Such linear-scatter behavior will lead to scattered light which remains in a coherent state as anticipated above for CRSS and further discussed below.

Nonlinearity of the spin dipole is revealed by considering the observable $\hat{J}_z$. The moments of $\hat{J}_z$ can be estimated directly from the distribution $|a_m|^2\propto e^{2 f_m}$ from Eq. (\ref{coeff}) and Fig. 2b. To this end, we expand the integral approximation of $f_m$ around its peak at $m=m_-$, and find for $j\gg 1$ that $|a_m|^2$ is well-described by a Gaussian,
\begin{eqnarray}
|a_m|^2\approx \frac{1}{\sqrt{2\pi w^2}}e^{-\frac{(m-m_-)^2}{2w^2}}, \quad  m_-\approx-j\sqrt{1-r^2},
\label{am}
\end{eqnarray}
of width $w^2=j r^2/(2\sqrt{1-r^2})$ (Appendix). This yields $\langle \hat{J_z}\rangle\approx m_-\approx-j\sqrt{1-r^2}$ in agreement with the mean-field solution of  superradiance \cite{DRUMMOND1978160,yelin,HL}, and exhibiting the second order phase transition at $r=1$ as discussed above. Together with $\langle \hat{J}_x-i\hat{J}_y\rangle=\langle \hat{J}\rangle=\alpha=jr e^{-i\varphi}$ and writing $r=\sin\theta$, this yields a spin vector $\hat{\mathbf{J}}=(\hat{J}_x,\hat{J}_y,\hat{J}_z)$ whose mean of length $j=N/2$ is directed at an angle $\theta$ away from the south pole of the Bloch sphere with an angle $\varphi$ along the $xy$ plane (Fig. 1b). This mean spin is equivalent to that of a CSS (Fig. 1a), however, the quantum fluctuations around it are different, as can be characterized by spin squeezing.

The spin squeezing parameter is defined by $\xi^2=(2j/|\langle\hat{\mathbf{J}}\rangle|^2)\min_{\phi}\mathrm{Var}[\hat{J}^{\bot}_{\phi}]$, where $\hat{J}^{\bot}_{\phi}$ is the projection of the spin vector onto an angle $\phi$ on the plane perpendicular to the direction of the mean spin $\langle\hat{\mathbf{J}}\rangle$ \cite{spinsqueezingreview,spinsqueezingparameter1,KIT}. $\xi^2<1$ signifies a non-classical spin-squeezed state wherein the constituent $N=2j$ spins-$1/2$ are pairwise entangled, and the best (minimum) achievable squeezing is bounded by the Heisenberg limit $\xi^2=1/N$ \cite{SOR,LEW,spinsqueezingreview}. Using property (\ref{eig}) of the CRSS, together with
$\langle\hat{J}_z\rangle= m_-$, $\mathrm{Var}[\hat{J}_z]=w^2$ from (\ref{am}), we find
$\xi^2=\sqrt{1-r^2}$ (Appendix). This result agrees with that obtained for steady-state superradiance using a mean-field approach (with linearized fluctuations) \cite{yelin,HL}, and as such, it is strictly exact only for $j\rightarrow \infty$. However, our CRSS theory allows to extend the analysis also for a finite $j\gg 1$. For a given $j$ we recall the validity region $0<r<r_j$ defined in Eq. (\ref{rj1}). Since minimal $\xi^2=\sqrt{1-r^2}$ is achieved for maximal $r$, then for a given $j$ the predictable optimal squeezing for a CRSS is given by
 \begin{subequations}
   \begin{tabularx}{\hsize}{@{}XXX@{}}
     \begin{equation}
        \xi^2_{\min}(j)= \sqrt{1-r^2_{j}},
        \label{xi1}
     \end{equation} &
     \begin{equation}
        \xi^2_{\min}(j)\sim \left(\frac{3}{2j}\right)^{\frac{1}{3}}\propto N^{-\frac{1}{3}}.
        \label{xi2}
     \end{equation}
   \end{tabularx}
 \end{subequations}
Here (\ref{xi2}) is obtained using $r_j$ from (\ref{rj2}) and expresses the asymptotic scaling of the CRSS optimal squeezing with $j=N/2\rightarrow \infty$. The predicted scaling $N^{-\frac{1}{3}}$ does not reach the Heisenberg limit $N^{-1}$.

\begin{figure}
    \centering
\includegraphics[width=\columnwidth]{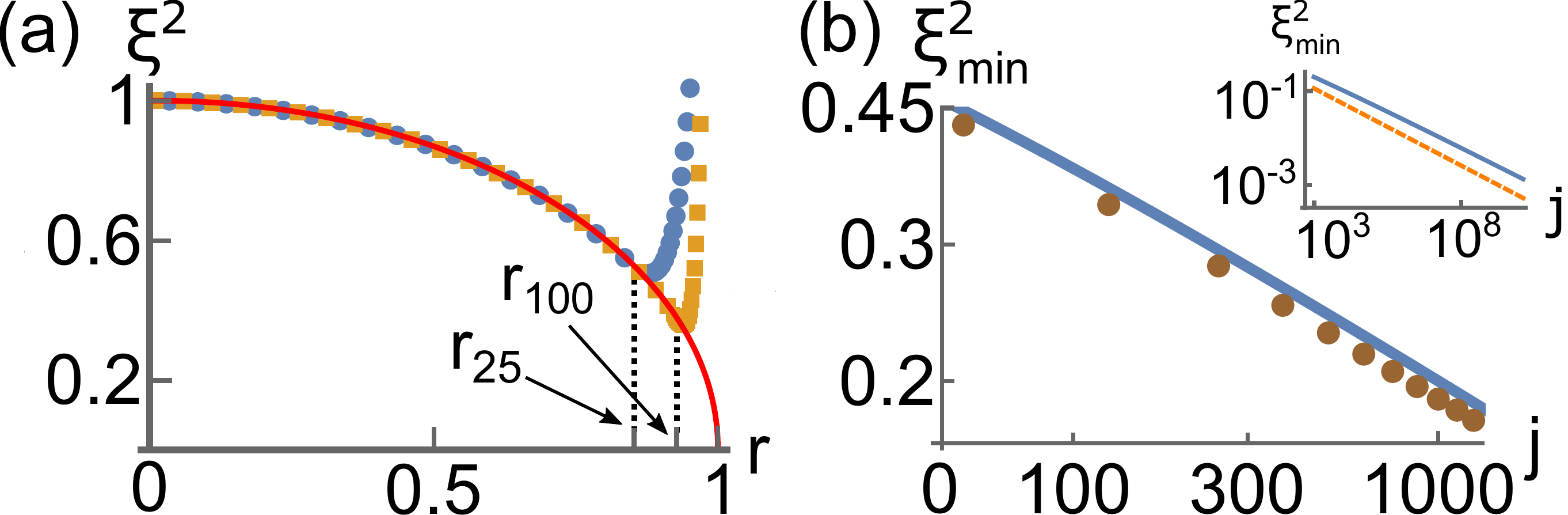}
    \caption{
    Spin squeezing $\xi^2$ in superradiance. (a) $\xi^2$ as a function of $r=|\Omega|/\Omega_c$ for the steady state of Eq. (\ref{ME}) with $j=N/2=25,100$ (dots, squares). The optimal (minimum) value $\xi_{\min}^2$ is seen around $r=r_j$ from Eq. (\ref{rj1}) [calculated using (\ref{eps})]. Agreement with CRSS theory (solid line) is observed within the validity region $r<r_j$. (b) $\xi_{\min}^2$ defined in (a) plotted as a function of $j$ in log-log scale (dots). Agreement with CRSS theory (\ref{xi1}) (solid line) is observed. Inset: increasing $j$, (\ref{xi1}) scales as $j^{-0.32}$, in close agreement with the asymptotics $j^{-\frac{1}{3}}$ from (\ref{xi2}) (solid and dashed lines, respectively).
    }
    \label{fig4}
\end{figure}

We now show how this theory applies also for spin squeezing in superradiance. In Fig. 4a we plot $\xi^2$ as a function of $r=|\Omega|/\Omega_c$ for the exact steady-state solution of Eq. (\ref{ME}) with $j=25,100$ (see \cite{SM} for calculation details, in analogy to \cite{Alejandro,yelin}). We observe excellent agreement with the theoretical CRSS result $\sqrt{1-r^2}$ within the expected validity region $r<r_j$ defined in (\ref{rj1}). Moreover, for $r>r_j$, $\xi^2$ begins to grow, setting its minimum at around $r=r_j$. This suggests that the CRSS prediction $\xi^2_{\mathrm{min}}$ from Eq. (\ref{xi1}) also predicts the best achievable (minimal) spin squeezing in superradiance. This is indeed what is observed in Fig. 4b, where the minimal $\xi^2$ of superradiance, obtained with respect to $r$ for different $j$ values as in Fig. 4a, is plotted as a function of $j$. Excellent agreement with the CRSS prediction from (\ref{xi1}) is exhibited. Within this range of $j$ values, a linear fit to the log-log plot yields the scaling $j^{-0.28}$. Extending the range of $j$ values as in the inset of Fig. 4b, $\xi^2_{\min}$ from Eq. (\ref{xi1}) scales as $j^{-0.32}$, already in close agreement with the predicted asymptotic scaling $j^{-\frac{1}{3}}$ from Eq. (\ref{xi2}).

\noindent\\
\textbf{Radiated light}
\\
After discussing the spin system and its observables, we now turn our attention to the light it radiates.
We begin by showing the counter-intuitive result, that the steady-state light emission in Dicke superradiance is a classical coherent state whose amplitude is linear with the incident driving field. This is identical to scattering by a linear dipole although the spin-$j$ dipole is highly nonlinear, and nonlinear optical systems typically generate photon interactions and quantum correlated light \cite{QNLOr}. The crucial point however is that we showed that in steady-state the superradiant spin is pumped into a CRSS, which is expected to radiate classically as argued above. To illustrate this, consider a generic relation between the field $\hat{E}$ in some detection mode and the radiating dipole $\hat{J}$,
\begin{eqnarray}
\hat{E}(t)=\hat{E}_{0}(t)+G\hat{J}(t),
\label{E}
\end{eqnarray}
written in the Heisenberg picture, similar to an input-output relation derivable from (\ref{H0}) \cite{MW,WEL}. Here $\hat{E}_{0}(t)$ is the freely evolving field, excluding interaction with the dipole, and the c-number $G$ is related to $\gamma$ (see Appendix, e.g. for superradiance in a cavity). In the Schr\"{o}dinger picture, the solution of Eq. (\ref{ME}) at a steady state time $t$ is a pure CRSS state of the dipole, $|j,\alpha\rangle_{d}$, and since the total dipole+field system is closed, then the total state is separable, $|\psi(t)\rangle=|j,\alpha\rangle_{d}\otimes |\chi\rangle_{f}$ with some field state $|\chi\rangle_{f}$. We now show that this state is a coherent state, i.e. an eigenstate of the (Schr\"{o}dinger picture) field operator $\hat{E}(0)$. Assuming a vacuum (or coherent) state for the field mode $\hat{E}(0)$ at initial time $t=0$, and using Eqs. (\ref{E}) and (\ref{eig}), we find $\hat{E}(0)|\psi(t)\rangle=G\hat{J}(0)|\psi(t)\rangle=G\alpha|\psi(t)\rangle$ (Appendix). This proves that the steady state radiation is a coherent state with an amplitude $G\alpha\propto \Omega$ linear in the incident field. So the nonlinear spin system driven to an entangled many-body state (CRSS), scatters light as a classical linear system. A similar relation (\ref{E}) also holds for the instantaneous emission at a given time $t$ without residing to steady-state conditions (Appendix), leading to the general conclusion that a CRSS radiates
coherent-state light.

\noindent\\
\textbf{Dipole-projected squeezing}
\\
Although the CRSS is a non-classical spin squeezed state, we found that it does not radiate quantum states such as squeezed light.
This is also intuitively understood as follows. From Hamiltonian (\ref{H0}) we observe that the field perceives the spin directly only through the dipole lowering operator $\hat{J}=\hat{J}_x-i\hat{J}_y$ and hence only through its projection onto the $xy$ ``dipole plane". This suggests that non-classical squeezed light is produced only if a similar squeezing exists in the projected $xy$ spin, hence motivating the distinction between spin squeezing of the total spin and that of the dipole-projected spin. To this end, we define the dipole quadrature $\hat{J}_{\phi}=(e^{i\phi}\hat{J}+e^{-i\phi}\hat{J}^{\dag})/2$, in analogy to the field quadrature $\hat{E}_{\phi}=e^{i\phi}\hat{E}+e^{-i\phi}\hat{E}^{\dag}$. Using Eq. (\ref{E}) with $[\hat{E},\hat{E}^{\dag}]=1$ and $[\hat{J},\hat{J}^{\dag}]=-2\hat{J}_z$, the relation between the variances of these quadratures is found
\begin{eqnarray}
\mathrm{Var}[\hat{E}_{\phi}]=1+4G^2 \left(\mathrm{Var}[\hat{J}_{\phi}]+\frac{1}{2}\langle\hat{J}_z\rangle\right),
\label{var}
\end{eqnarray}
where $G$ is taken real without loss of generality. Considering the Heisenberg uncertainty
$\mathrm{Var}[\hat{J}_{\phi}]\mathrm{Var}[\hat{J}_{\phi+\frac{\pi}{2}}]\geq |\langle\hat{J}_z\rangle|^2/4$, we define the \emph{dipole-projected squeezing} for a quadrature $\phi$ as $\mathrm{Var}[\hat{J}_{\phi}]<|\langle\hat{J}_z\rangle|/2$. Then, from Eq. (\ref{var}) we conclude that for $\langle\hat{J}_z\rangle<0$ the emitted light becomes squeezed, i.e. $\mathrm{Var}[\hat{E}_{\phi}]<1$, provided that the dipole-projected spin is also squeezed.

For CRSS, using the definition (\ref{eig}) we obtain $\mathrm{Var}[\hat{J}_{\phi}]=-\langle\hat{J}_z\rangle/2$, so that no dipole squeezing exists and vacuum noise-level is obtained for the light, $\mathrm{Var}[\hat{E}_{\phi}]=1$, as expected for a coherent state. This is nicely seen by the geometrical picture of Fig. 1b: the noise contour perpendicular to the mean spin direction $\theta, \varphi$ presents the spin squeezing of CRSS with the $\varphi$ and $\varphi+\frac{\pi}{2}$ axes exhibiting increased and reduced variances $(j/2)/\cos\theta$ and $(j/2)\cos\theta$, respectively (Appendix). However, upon projection of the noise contours onto the $xy$ dipole plane, the $\varphi$ axis variance is multiplied by a factor $\cos^2\theta$ so that both axes have identical noises and no projected squeezing exists.

For a CSS, the opposite situation occurs, as seen in Fig. 1a. Whereas the noise contour at the plane perpendicular to the main spin direction is a circle of variance $j/2$ so that no spin squeezing exists \cite{spinsqueezingreview}, the projection onto the $xy$ dipole plane reduces the noise at the $\varphi$ direction to $(j/2)\cos^2\theta$, smaller than $|\langle \hat{J}_z \rangle|/2=(j/2)\cos\theta$. This means that a CSS at the southern Bloch hemisphere, $\langle \hat{J}_z \rangle<0$, radiates squeezed light. The latter is also verified by a direct calculation of $\mathrm{Var}[\hat{E}_{\phi}]$ using a CSS \cite{SM}.

The above results can be summarized in the following counter-intuitive manner: CSS, which is a non-correlated many-body spin, emits quantum-correlated squeezed light; whereas CRSS, which is a quantum-correlated (spin-squeezed) many-body spin, emits non-correlated coherent light. The fact that correlations are necessary in a spin system to produce classical non-correlated light is nicely seen from a microscopical picture. For dipole-projected squeezing, and hence light squeezing to exist, the phase-dependent variance $\mathrm{Var}[\hat{J}]=\langle\hat{J}^2\rangle-\langle\hat{J}\rangle^2$ also has to exist. Writing this quantity using constituent spin operators $\hat{J}=\sum_{n=1}^N \hat{\sigma}_n$ and recalling $\hat{\sigma}_n^2=0$, we have
\begin{eqnarray}
\mathrm{Var}[\hat{J}]=-\sum_{n}\langle\hat{\sigma}_n\rangle^2+
\sum_n\sum_{n\neq m}\left(\langle\hat{\sigma}_n\hat{\sigma}_m\rangle-\langle\hat{\sigma}_n\rangle\langle\hat{\sigma}_m\rangle\right).
\label{J2}
\end{eqnarray}
Assuming a coherent drive $\Omega$, where $\langle\hat{\sigma}_n\rangle\neq 0$, and for statistically independent atoms as in a CSS, the second term vanishes such that $\mathrm{Var}[\hat{J}]\neq 0$ and light squeezing exists. This typical situation occurs in resonance fluorescence from an ensemble of individual atoms, where the nonlinearity of each two-level atom generates squeezed light \cite{singleatom,Natom}. For squeezing to vanish, we must have a non-vanishing second term in Eq. (\ref{J2}), namely, non-vanishing correlations between atoms. For a pure state, these correlations imply entanglement, which is exactly the situation in a CRSS, where the pairwise entanglement due to spin-squeezing leads to the exact cancellation of the two terms in (\ref{J2}).

\noindent\\
\textbf{Outlook}
\\
This study introduces CRSS as a new class of macroscopic spin states which possess rather unusual properties: they are many-body entangled states which behave as macroscopic classical emitters.

These new insights open several important directions.
First, since CRSS theory was shown here to describe Dicke superradiance, the above predictions on both photon and collective-spin observables could be evaluated experimentally in platforms such as cavities \cite{HAR,TOMs1}, waveguides
\cite{Alejandro,KIMsr,CHArmp}, and superconducting resonators \cite{MAJ}. Moreover, CRSS theory may be conceptually and technically useful for the study of various related collective radiation phenomena. For example, it is interesting to explore the possible role of CRSS in superradiant lasing \cite{HOL,TOM1} and whether it can be generalized to account for collective radiation beyond the permutation-symmetric Dicke case \cite{GROSS1982301,YEL}. This could have important consequences for understanding and exploiting superradiance both in ordered and disordered ensembles \cite{coop,RUI,ANA,RUO3,KSR1,BRW2,YEb} and for quantum metrology \cite{CHAYE,HEN,REY,YEr}.

Another direction is the exploration of the very nature of CRSS and its relation to many-body entangled spin states. The optimal spin squeezing contained in a CRSS exhibits a different scaling, $N^{-\frac{1}{3}}$, than that of spin-squeezed states generated by one- and two-axis twisting Hamiltonians \cite{spinsqueezingreview,KIT}, with $N^{-\frac{2}{3}}$ and $N^{-1}$, respectively. It is interesting to study the existence of unitary transformations that generate this different class of spin squeezing. Moreover, although CRSS underlies the dissipative Dicke phase transition, it is in fact a pure state. Hence, it might be possible to relate it to quantum phase transitions in Hamiltonian many-body spin systems.

More generally, relying on the finding that a macroscopic spin is perceived as a classical emitter provided that its constituents are quantum entangled, it should be considered if and how this idea extends to other composite quantum systems.

\section*{APPENDIX}
\noindent\\
\textbf{Radiation from a lowering-operator eigenstate}
\\
Here we show that for a generic dipole+field system, $\hat{H}=\hbar g(\hat{E}\hat{d}^{\dag}+\hat{E}^{\dag}\hat{d})$ (coupling constant $g$), that is pumped into an eigenstate of $\hat{d}$ with an eigenvalue $\alpha$, one obtains a coherent state of the field $\hat{E}$, which is equivalent to that obtained by the replacement $\hat{d}\rightarrow \alpha$ in $\hat{H}$. We assume that the system is pumped to a state $|\psi(t)\rangle$ that satisfies at any given time $t$,
$\hat{d}(0)|\psi(t)\rangle=\alpha |\psi(t)\rangle$, where $\hat{d}(0)$ is the Scr\"{o}dinger picture operator. Such a pumping mechanism can be supplied by an external reservoir as in the steady state of Eq. (\ref{ME}) with $\hat{J} \leftrightarrow \hat{d}$ and as discussed above for superradiance.
We note
\begin{eqnarray}
\hat{E}(0)|\psi(t)\rangle=\hat{U}(t)\hat{E}(t)|\psi(0)\rangle, \quad \hat{U}(t)=e^{-\frac{i}{\hbar}\hat{H}t},
\label{s1}
\end{eqnarray}
where $|\psi(t)\rangle=\hat{U}(t)|\psi(0)\rangle$ and $\hat{E}(t)=\hat{U}^{\dag}(t)\hat{E}(0)\hat{U}(t)$ is the Heisenberg picture operator, formally obtained from the Heisenberg equations as
\begin{eqnarray}
\hat{E}(t)=\hat{E}(0)-ig\int_0^tdt'\hat{d}(t').
\label{s2}
\end{eqnarray}
Inserting Eq. (\ref{s2}) into Eq. (\ref{s1}), assuming an initial vacuum state of the field, $\hat{E}(0)|\psi(0)\rangle=0$, and using $\hat{d}(t')=\hat{U}^{\dag}(t')\hat{d}(0)\hat{U}(t')$, we obtain
\begin{eqnarray}
&&\hat{E}(0)|\psi(t)\rangle=-ig\hat{U}(t)\int_0^t dt'\hat{U}^{\dag}(t')\hat{d}(0)|\psi(t')\rangle
\nonumber\\
&&=-ig\alpha\hat{U}(t)\int_0^t dt' \hat{U}^{\dag}(t')|\psi(t')\rangle=-ig\alpha t|\psi(t)\rangle,
\label{s3}
\end{eqnarray}
where the pumping assumption, $\hat{d}(0)|\psi(t')\rangle=\alpha |\psi(t')\rangle$, was used in the second equality. Equation (\ref{s3}) then shows that $|\psi(t)\rangle$ is a coherent state with an amplitude $-ig\alpha t$ just as one would get from the displacement operator $\hat{U}(t)=e^{-\frac{i}{\hbar}\hat{H}t}$ formed by setting $\hat{d}\rightarrow \alpha$ in $\hat{H}$.
\\
\noindent\\
\textbf{Minimization of the proximity error}
\\
We showed that the error $\epsilon_{j}(r)$ of state (\ref{ans}) goes to zero exponentially. Here we show that it also does so at the highest rate. We define the state $\ket{\alpha,j}_{\mathrm{min}}$ that minimizes
$\epsilon_{j}(r)$ in Eq. (\ref{limit}) while keeping $j$ and $\alpha$ fixed. Writing the error for a state $\ket{\psi}$ as $\epsilon_{\psi}=\left\Vert \hat{K}|\psi\rangle\right\Vert$ with $\hat{K}=\hat{J}-\alpha$, $\ket{\alpha,j}_{\mathrm{min}}$ is given by the state $\ket{\psi}$ which minimizes $\epsilon^2_{\psi}=\bra{\psi}\hat{K}^{\dag}\hat{K}\ket{\psi}$. Since the lower bound of this quantity is the smallest eigenvalue of $\hat{K}^{\dag}\hat{K}$, its corresponding eigenstate is in fact the state $\ket{\alpha,j}_{\mathrm{min}}$, and can be readily evaluated numerically for given $j$ and $\alpha$. Figure 5a compares the two definitions $\ket{j,\alpha}_{\mathrm{ans}}$ and $\ket{j,\alpha}_{\mathrm{min}}$, by displaying the infidelity $1-|_{\mathrm{ans}}\langle j,\alpha| j,\alpha\rangle_{\mathrm{min}}|$. It is seen that this infidelity decreases exponentially with $j$, indicating that the two definitions coalesce in the relevant limit $j\rightarrow\infty$.
\\
\noindent\\
\textbf{Superradiance at finite $j$}
\\
For a finite $j\gg 1$, the state $\ket{j,\alpha}_{\mathrm{ans}}$ from (\ref{ans}) is a CRSS to an excellent approximation in the range $r<r_j$ with $r_j$ from (\ref{rj1}). We thus expect that within this range $\ket{j,\alpha}_{\mathrm{ans}}$ also well approximates the steady-state of the superradiance master equation (\ref{ME}). For the case $\Delta=0$, the latter is given by $\hat{\rho}_s\propto[(\hat{J}-\alpha)^{\dag}(\hat{J}-\alpha)]^{-1}$ \cite{CAR,LAW} and can be readily evaluated numerically for given $j$ and $r=|\Omega|/\Omega_c$. In Fig. 5b we plot the infidelity between the state (\ref{ans}) and the exact steady-state $\hat{\rho}_s$,
$jr(1-_{\mathrm{ans}}\bra{j,\alpha}\hat{\rho}_s\ket{j,\alpha}_{\mathrm{ans}})$, observing a similar dependence on $j$ and $r$ as that of the CRSS proximity error in Fig. 3b. In particular, the curve $r_j$ (dashed line) indeed overlaps with an infidelity contour beyond which the state $\ket{j,\alpha}_{\mathrm{ans}}$ well approximates $\hat{\rho_s}$.
\begin{figure}
    \centering
    \includegraphics[width=\columnwidth]{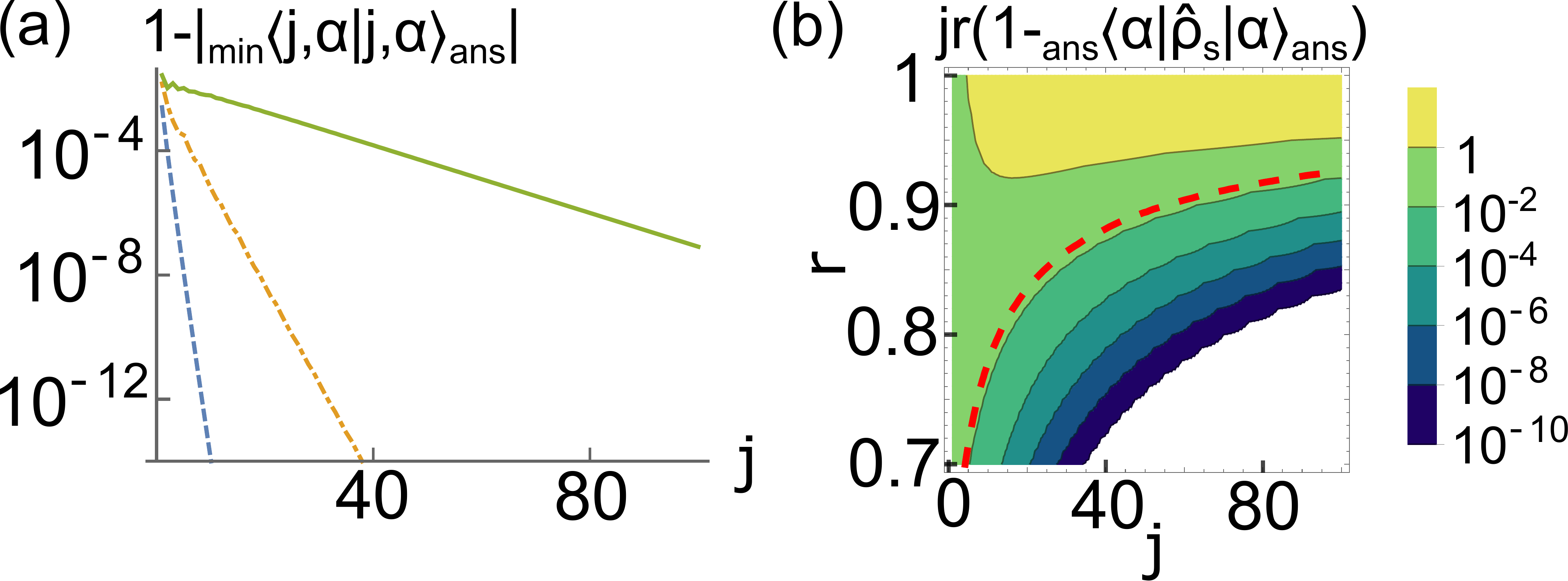}
    \caption{
    (a) The infidelity between the state $\ket{j,\alpha}_{\min}$ that minimizes the proximity error and the CRSS ansatz $\ket{j,\alpha}_{\mathrm{ans}}$ decays exponentially with $j$, so that the states coalesce for $j\rightarrow \infty$ ($r=0.4,0.7,0.9$ in dashed, dash-dotted, and solid lines). (b) The infidelity between the exact steady-state $\hat{\rho}_s$ of (\ref{ME}) and the CRSS ansatz behave similar to the proximity error in Fig. 3b, as expected. Therefore, $\hat{\rho}_s\approx \ket{j,\alpha}_{\mathrm{ans}}\bra{j,\alpha}$ forms an excellent approximation at any finite $j\gg 1$ for $0<r<r_j$ (dashed curve marks $r_j$, white parts represent values too small for our numerical evaluation).
    }
    \label{figM}
\end{figure}
The normalization factor $jr$ in the above infidelity definition is analogous to that in the proximity error in Eqs. (\ref{limit}) and (\ref{ans}). This error (or infidelity) definition means that an error of order $<1$ leads to values of typical spin correlators (e.g. $\langle\hat{J}^2\rangle$) which agree with their CRSS values ($\langle\hat{J}\rangle^2$) up to a correction smaller by a factor $\sim 1/|\alpha|=1/(jr)$ (similar to a mean-field assumption).
\\\noindent\\
\textbf{Expansion of the distribution $|a_m|^2$}
\\
From Eq. (\ref{coeff}) we have $|a_m|^2\propto e^{2f_m}$. Approximating the sum in $f_m$ by an integral \cite{SM}, we obtain an analytical expression for $f_m=f(m)$ and expand it around the peak $m=m_-$,
\begin{eqnarray}
f(m)&\approx& A_0-A_2(m-m_-)^2
\nonumber\\
&+&A_3(m-m_-)^3+A_4(m-m_-)^4,
\label{s6}
\end{eqnarray}
with the expansion coefficients for $j\gg 1$,
\begin{eqnarray}
A_2(j,r)&\approx& \frac{\sqrt{1-r^2}}{2r^2}\frac{1}{j},
\nonumber\\
A_3(j,r)&\approx& \left(\frac{1}{3r^4}-\frac{1}{6r^2}\right)\frac{1}{j^2},
\nonumber\\
A_4(j,r)&\approx& \frac{\sqrt{1-r^2}}{3r^4}\left(\frac{1}{4}-\frac{1}{r^2}\right)\frac{1}{j^3},
\label{s7}
\end{eqnarray}
and where $A_0$ is of no importance here. For a Gaussian approximation to be valid, only the $A_2$ term should be significant. This term implies a distribution width of order
$w=\left[j r^2/(2\sqrt{1-r^2})\right]^{\frac{1}{2}}\propto \sqrt{j}$, so that the terms $A_n\propto j^{-n+1}$ for $n>2$ are negligible if  $A_n w^n\ll 1$. It can be verified that this is indeed the case if we demand $\sqrt{1-r^2}j^{\frac{1}{3}}\gg 1$. This is equivalent to the CRSS validity condition $r<r_j$ with the asymptotic $r_j$ from Eq. (\ref{rj2}). In this case, we obtain a Gaussian distribution and, demanding normalization, we arrive at Eq. (\ref{am}).
\\\noindent\\
\textbf{CRSS spin squeezing calculation}
\\
For the spin squeezing, one has to calculate the variance of the spin operator $\hat{J}^{\bot}_{\phi}$ that is projected to the plane perpendicular to the mean spin vector (MSV). The latter defines the angles $\theta$ and $\varphi$ from the main text, and hence the corresponding rotated coordinate system in which the MSV points to a rotated $z'$ axis. Within this rotated coordinate system, the lowering spin operator $\hat{J}'=\hat{J}'_x-i\hat{J}'_y$ (with $\hat{J}'_{i}$ the spin projection onto the rotated $i\in\{x',y'\}$ axis) is given in terms of the original spin operators via the transformation,
\begin{eqnarray}
\hat{J}'=e^{i\varphi}\frac{\cos{\theta}+1}{2}\hat{J}+e^{-i\varphi}\frac{\cos{\theta}-1}{2}\hat{J}^{\dag}-\sin\theta\hat{J}_z,
\label{s8}
\end{eqnarray}
recalling $\sin\theta=r$. A general Hermitian spin operator on the plane perpendicular to the MSV is then given by $\hat{J}^{\bot}_{\phi}=(e^{i\phi}\hat{J}'+e^{-i\phi}\hat{J}'^{\dag})/2$. Calculating the variance of this operator involves various first and second order moments of $\hat{J}$,  $\hat{J}^{\dag}$ and $\hat{J}_z$, all of which can be evaluated using the CRSS property (\ref{eig}), the moments of $\hat{J}_z$ from Eq. (\ref{am}) and the SU(2) commutation relations $[\hat{J},\hat{J}^{\dag}]=-2\hat{J}_z$ and $[\hat{J},\hat{J}_z]=\hat{J}$. Recalling the mean spin length $|\langle\hat{\mathbf{J}}\rangle|=j$ in a CRSS, we finally have
\begin{eqnarray}
\xi^2_{\phi}=\frac{2j \mathrm{Var}[\hat{J}^{\bot}_{\phi}]}{|\langle\hat{\mathbf{J}}\rangle|^2}=\cos[2(\phi-\varphi)]\frac{1-c^2}{2c}+\frac{1+c^2}{2c},
\label{s9}
\end{eqnarray}
with $c=\sqrt{1-r^2}$.
The minimal noise is obtained for $\phi=\varphi+\pi/2$, yielding the spin squeezing $\xi^2=\xi^2_{\varphi+\pi/2}=c=\sqrt{1-r^2}=\cos\theta<1$, whereas the conjugate quadrature $\phi=\varphi$ exhibits anti-squeezing, $\xi^2_{\varphi}=1/\cos\theta>1$. This is illustrated in Fig. 1b.
\\
\noindent\\
\textbf{The scattered field}
\\
The relation (\ref{E}) between the output field and the radiating dipole can be generally obtained in the Heisenberg picture. One typical case is that of a dipole damped by a continuum of photon modes described by an operator $\hat{E}$ in (\ref{H0}), in equivalence to the master equation (\ref{ME}) and in analogy to input-output theory \cite{MW} (where $G$ is often related to the photon Green's function \cite{WEL,CARb}). For example, consider superradiance in a leaky cavity, where the atoms are situated at positions where they are identically coupled to the cavity mode $\hat{a}$ and driven by a laser $\Omega$. The Hamiltonian is given by
\begin{eqnarray}
\hat{H}=-\hbar\delta_a\hat{J}_z-\hbar\delta_c\hat{a}^{\dag}\hat{a}+\hbar\left[\left(g^{\ast}\hat{a}^{\dag}+\Omega^{\ast}\right)\hat{J}+\mathrm{h.c.}\right],
\label{s4}
\end{eqnarray}
where $\delta_a$ and $\delta_c$ are the detunings of the dipole and the cavity from the driving-laser frequency. Considering the damping of the cavity to the outside modes thorough its leaky mirrors at a rate $\kappa$, and exploiting the separation of time scales $\kappa\gg g$, we formally solve the Heisenberg-Langevin equation for $\hat{a}$ at $t\gg 1/\kappa$, obtaining
\begin{eqnarray}
g\hat{a}(t)=\left(\Delta-i\frac{\gamma}{2}\right)\hat{J}(t)+\hat{f}(t), \quad \Delta-i\frac{\gamma}{2}=\frac{|g|^2}{\delta_c+i\kappa/2}.
\label{s5}
\end{eqnarray}
Here $\hat{f}(t)$ is a Langevin noise due to the freely evolving (vacuum) field of the modes to which the cavity is damped, satisfying $\hat{f}(t)|\psi(0)\rangle=0$, where the initial state $|\psi(0)\rangle$ is in the field's vacuum.
We identify that Eq. (\ref{s5}) is equivalent to Eq. (\ref{E}) with the ``detection mode" $\hat{E}$ being the cavity mode. A similar relation can be obtained between the dipole and the outside field modes. Once the relation (\ref{E}) is established, and using $\hat{E}_{0}(t)|\psi(0)\rangle=0$ for the freely evolving vacuum, a similar algebra to that of Eq. (\ref{s3}) leads to
$\hat{E}(0)|\psi(t)\rangle=G\hat{U}(t)\hat{J}(t)|\psi(0)\rangle=G\hat{J}(0)|\psi(t)\rangle=G\alpha|\psi(t)\rangle$.

A simpler case wherein a relation of the type (\ref{E}) holds is that of the instantaneous field emitted at a given time $t\equiv 0$. The general solution from Eq. (\ref{s2}), for a dipole $\hat{d}=\hat{J}$ and a short evolution time $\tau$ after $t=0$, is given by $\hat{E}(\tau)\approx \hat{E}(0)-ig\tau \hat{J}(0)$. Here again it is simple to show that $\hat{E}(0)|\psi(\tau)\rangle=G\alpha|\psi(\tau)\rangle$ with $G=-ig\tau$.

\begin{acknowledgments}
We acknowledge financial support from the Israel Science Foundation (ISF) grant No. 2258/20, the ISF and the Directorate for Defense Research and Development (DDR\&D) grant No. 3491/21, the Center for New Scientists at the Weizmann Institute of Science, the Council for Higher Education (Israel), and QUANTERA (PACE-IN).
This research is made possible in part by the historic generosity of the Harold Perlman Family.
\end{acknowledgments}

\bibliographystyle{ieeetr}
\bibliography{bibfile_np}

\end{document}